\documentclass[12pt,a4paper]{article}

\usepackage{amsmath}
\usepackage{amssymb}
\usepackage{graphicx}
\usepackage{dsfont}
\usepackage{overpic}
\usepackage{cite}
\usepackage{slashed}

\let\oldappendix=\appendix
\let\oldsection=\section
\renewcommand{\appendix}{\oldappendix%
\def\theequation{\Alph{section}.\arabic{equation}}%
\renewcommand{\section}{\setcounter{equation}{0}\oldsection}}

\newcommand{\beq}{\begin{equation}}
\newcommand{\eeq}{\end{equation}}
\newcommand{\beqa}{\begin{eqnarray}}
\newcommand{\eeqa}{\end{eqnarray}}
\newcommand{\no}{\nonumber}

\newcommand{\tr}{\mbox{tr}}

\def\bra#1{\left\langle #1\right|}
\def\ket#1{\left| #1\right\rangle}

\newcommand{\newop}[2]{\def#1{\mathop{\mathrm{#2}}\nolimits}}
\newop{\artanh}{artanh}
\newop{\det}{det}
\newop{\tr}{tr}
\newop{\diag}{diag}
\newop{\Re}{Re}
\newop{\Im}{Im}

\newcommand{\Lagr}{\mathcal{L}}

\begin{document}

\hfill 

\hfill 

\bigskip\bigskip

\begin{center}

{{\Large\bf  Gauge invariance in two-particle scattering\footnote{Work supported in part by the DFG
through funds provided to the  TR 16 ``Subnuclear structure of matter''.}}}

\end{center}

\vspace{.4in}

\begin{center}
{\large B.~Borasoy\footnote{email: borasoy@itkp.uni-bonn.de}$^a$,
        P.~C.~Bruns\footnote{email: bruns@itkp.uni-bonn.de}$^a$,
        U.-G.~Mei{\ss}ner\footnote{email: meissner@itkp.uni-bonn.de}$^{a,b}$,
        R.~Ni{\ss}ler\footnote{email: rnissler@itkp.uni-bonn.de}$^a$}

\bigskip

\bigskip

$^a$Helmholtz-Institut f\"ur Strahlen- und Kernphysik (Theorie) \\
Universit\"at Bonn, Nu{\ss}allee 14-16, D-53115 Bonn, Germany \\[0.3cm]
$^b$Institut f\"ur Kernphysik (Theorie), Forschungszentrum J\"ulich \\
D-52425 J\"ulich, Germany \\

\vspace{.2in}

\end{center}

\vspace{.7in}

\thispagestyle{empty} 

\begin{abstract}
It is shown how gauge invariance is obtained for the coupling of 
a photon to a two-body state described by the
solution of the Bethe-Salpeter equation. This is illustrated both for a complex scalar field theory
and for interaction kernels derived from chiral effective Lagrangians.
\end{abstract}\bigskip

\begin{center}
\begin{tabular}{ll}
\textbf{PACS:}& 11.10.St, 12.39.Fe \\[6pt]
\textbf{Keywords:}& Bethe-Salpeter equation, gauge invariance, chiral Lagrangians.
\end{tabular}
\end{center}

% 11.30.Rd Chiral symmetries
% 12.39.Fe Chiral Lagrangians
% 13.75.Gx Pion-baryon interactions
% 13.75.Jz Kaon-baryon interactions
% 11.80.-m Relativistic scattering theory
% 11.80.Gw Multichannel scattering
% 11.10.St Bethe-Salpeter equations

\vfill

%%%%%%%%%%%%%%%%%%%%%%%%%%%%%%%%%%%%%%%%%%%%%%%%%%%%%%%%%%%%%%%%%%%%%%%%%%%%%%%%
%%%%%%%%%%%%%%%%%%%%%%%%%%%%%%%%%%%%%%%%%%%%%%%%%%%%%%%%%%%%%%%%%%%%%%%%%%%%%%%%
\section{Introduction}\label{sec:Intro}
%%%%%%%%%%%%%%%%%%%%%%%%%%%%%%%%%%%%%%%%%%%%%%%%%%%%%%%%%%%%%%%%%%%%%%%%%%%%%%%%

Chiral perturbation theory provides an appropriate framework for studying
hadronic processes at low energies \cite{GL}. In chiral perturbation theory (ChPT)
the most general effective Lagrangian 
incorporating the same symmetries and symmetry breaking patterns as the underlying
theory, QCD, is formulated in terms of the relevant degrees of freedom, i.e.\ mesons and baryons. 
Moreover, external fields representing, e.g., 
external photons are included in a gauge invariant fashion.
The Green's functions are then expanded perturbatively in powers of Goldstone boson masses 
and small three-momenta. By employing a regularization scheme which respects
chiral symmetry, gauge invariance is maintained at every order in the loop
expansion of ChPT.

However, the systematic loop expansion involves a
characteristic scale $\Lambda_\chi \simeq 4\pi F_\pi \approx 1.2$~GeV at which the
chiral series is expected to break down and the limitation to very low-energy
processes is even enhanced in the vicinity of resonances. The appearance
of resonances in certain channels constitutes a major problem for the
loop expansion of ChPT, since a resonance cannot be reproduced
at any given order of the chiral series. Nevertheless, at low energies the
contribution from such resonances is encoded in the numerical values of
certain low-energy constants (frequently called resonance saturation).

Recently, considerable effort has been undertaken to combine the effective
chiral Lagrangian approach with non-perturbative methods, both
in the meson-baryon sector \cite{KSW,OM,LK} and in the purely mesonic
sector \cite{OO}.
The combination with non-perturbative
schemes has made it possible to go to energies beyond $\Lambda_\chi$
and to generate resonances dynamically (giving up, however, certain aspects of 
the rigorous framework constituted by ChPT). Two prominent examples
in the baryonic sector are the $\Lambda(1405)$ and the $S_{11}(1535)$.
The first one is an $s$-wave resonance just below the $K^- p$
threshold and dominates the interaction of the $\bar{K} N$ system, while
the $S_{11}(1535)$ was identified in \cite{KSW} 
as a quasi-bound $K \Lambda$-$K \Sigma$ state.

Such chiral unitary approaches have been extended to photo- and electroproduction processes
of mesons on baryons, see e.g. \cite{KWW, BWW, BMW}. In these coupled channel models
the initial photon scatters with the incoming baryon into a meson-baryon pair
which in turn rescatters (elastically or inelastically) an arbitrary number of times.
The two-body final state interactions are taken into account in a coupled-channels Bethe-Salpeter
equation (BSE) or -- in the non-relativistic framework -- Lippmann-Schwinger equation.
The coupling of the incoming photon to other possible vertices is omitted,
and although these approaches appear to describe the available data well, the issue
of gauge invariance is not discussed in these chiral unitary approaches.
On the other hand, a method to obtain conservation of the electromagnetic 
current of a two-nucleon system
is presented in \cite{GR} and extended to a resonance model for pion photoproduction in \cite{SG}.
Alternatively, the so-called ``gauging of equations'' method has been developed
in \cite{KB1, KB2} to incorporate an external electromagnetic field 
in the integral equation of a few-body system in a gauge invariant fashion.
Further investigations of gauge invariance in pion photoproduction within $\pi N$ models
can be found in \cite{AA,H}.
Gauge invariance is also of interest in coupled-channels approaches in the mesonic sector,
e.g in radiative $\phi, \rho$ decays \cite{Ol, MHOT} and in the anomalous
decays $\eta, \eta' \to \gamma \gamma, \pi^+ \pi^- \gamma$ \cite{BN1, BN2}.
For related recent work, see also \cite{HN}.

The purpose of the present work is to illustrate 
how gauge invariance can be maintained when an external photon
couples to a two-particle state described by the BSE within the chiral effective framework.
We will start in the next section by first outlining the procedure for a simple scalar field theory. 
With the insights gained from this example we can then address gauge
invariance in meson-baryon scattering processes with chiral effective Lagrangians.
In Sec.~\ref{sec:wt} we discuss the case with the interaction kernel of the BSE derived
from the leading order Lagrangian -- the Weinberg-Tomozawa term.
The extension to driving terms from higher chiral orders
is presented in Sec.~\ref{sec:nlo}.
In Section \ref{sec:uni} it is shown that the corresponding amplitudes also satisfy
unitarity constraints.
Our conclusions and outlook are presented in Sec.~\ref{sec:conclusions}.

%%%%%%%%%%%%%%%%%%%%%%%%%%%%%%%%%%%%%%%%%%%%%%%%%%%%%%%%%%%%%%%%%%%%%%%%%%%%%%%%
\section{Complex scalar fields} \label{sec:complex}

In this section, we will demonstrate in a simple field theory with complex
scalar fields that the coupling of an external photon to a
two-body state, which corresponds to the solution of the BSE, is gauge invariant.
To this end, consider the (normal-ordered) Lagrangian for complex fields $\phi$ and $\psi$
\beq  \label{Lagrpart}
\Lagr =  \partial_\mu \phi^* \partial^\mu \phi - m^2 \phi^* \phi 
        + \partial_\mu \psi^* \partial^\mu \psi - M^2 \psi^* \psi - g (\phi^* \phi)(\psi^* \psi)~,
\eeq
with masses $m$ and $M$, respectively.
For small values of the coupling constant $g$, the scattering process $\phi(p_1) \psi(p_2)
\to  \phi(p_3) \psi(p_4)$ may be calculated perturbatively. For general values
of $g$, however, and if one is interested in bound states, one must resort to
non-perturbative techniques such as the BSE. The Bethe-Salpeter equation for
the scattering matrix $T$ of the two-particle scattering process can be written
as
\beq \label{BSE}
T(s) = g + g G(s) T(s) \ ,
\eeq
where $s= p^2 = (p_1 +p_2)^2 = (p_3+p_4)^2$ and $G$ is the scalar loop integral
\beq \label{BSEint}
G(p^2) = i \int_l \  \Delta_\phi(l) \Delta_\psi(p+l)
\eeq
utilizing the short-hand notation
\beq
\int_l = \int \frac{d^4 l}{(2 \pi)^4} 
\eeq
and the propagators 
\beq
i \Delta_\phi(l) = \frac{i}{l^2 - m^2} \ , \qquad  i \Delta_\psi(l) = \frac{i}{l^2 - M^2}\ .
\eeq
The solution of the BSE is given by
\beq \label{BSEsol}
T(s) = \frac{g}{1 - g G(s)} = g + g G(s) g + g G(s) g G(s) g + \ldots  
\eeq
which amounts to the summation of an infinite series of $s$-channel loops, 
the so-called bubble chain (or bubble sum).
From inversion of Eq.~(\ref{BSEsol}) it immediately follows that
\beq
\Im T^{-1} = - \Im G =  \frac{|\mbox{\boldmath$q$}_{cm}|}{8 \pi \sqrt{s}} \ \theta (s - (m+M)^2)  \ ,
\eeq
where the second relation is deduced from unitarity of the amplitude $T$
and $\mbox{\boldmath$q$}_{cm}$ is the three-momentum in the center-of-mass frame.
The explicit calculation of the scalar loop integral $G$, e.g. in dimensional
regularization, indeed confirms the above relation.
%Strictly speaking, the integral $G$ is divergent, but its divergent pieces can be
%absorbed by renormalizing the coupling $g$. In the following, we will thus assume that
%the scalar loop integral $G$ has been rendered finite by an appropriate redefinition of the
%coupling and omit the divergent pieces henceforth.

Let us now turn to the coupling of an external photon field to the solution
of the BSE. By minimal substitution 
\beqa
\partial_\mu &\to& \nabla_\mu = \partial_\mu + i e_{\phi} {\cal A}_\mu \qquad \mbox{for} \ \phi \ , \no \\
\partial_\mu &\to& \nabla_\mu = \partial_\mu + i e_{\psi} {\cal A}_\mu \qquad \mbox{for} \ \psi \ ,
\eeqa
where ${\cal A}_\mu$ is the photon field and $e_{\phi} \, (e_{\psi} )$ the charge 
of the meson $\phi \, (\psi)$, one obtains a locally gauge invariant Lagrangian.
The coupling of the photon with incoming momentum $k$ to the four external
legs of a bubble chain leads to the amplitudes 
\begin{eqnarray}
T^{\mu}_{1} &=&  e_\phi \, T(s')\, \Delta(p_{1}+k) \,  (2p_{1}+k)^{\mu} \ , \no \\
T^{\mu}_{2} &=&  e_\psi \, T(s')\, \Delta(p_{2}+k) \,  (2p_{2}+k)^{\mu} \ , \no \\
T^{\mu}_{3} &=& e_\phi (2p_{3}-k)^{\mu} \, \Delta(p_{3}-k) \, T(s) \ , \no \\
T^{\mu}_{4} &=& e_\psi (2p_{4}-k)^{\mu} \, \Delta(p_{4}-k) \, T(s) \ ,
\end{eqnarray}
where $s' = (p+k)^2$.
Multiplying these contributions with the four-momentum of the photon, $k_\mu$,
and setting the external legs on-shell yields
\beq
( e_\phi + e_\psi) \,  \left[ T(s') - T(s) \right]~,
\eeq
which in general does not vanish. This underlines that, in order to achieve gauge invariance,
it is not sufficient to couple the photon only to external legs.
One rather has to include contributions which arise due to the coupling of
the photon to intermediate states within the bubble chain, 
leading to the additional contributions 
\beqa
T^{\mu}_{5} &=& i e_\phi T(s')  \int_l \ \Delta_\phi(l+k) \, (2l+k)^{\mu} \, \Delta_\phi(l) \, 
                 \Delta_\psi(p-l) \ T(s) \ , \no \\
T^{\mu}_{6} &=& i e_\psi T(s')  \int_l \ \Delta_\psi(l+k) \, (2l+k)^{\mu} \, \Delta_\psi(l) \, 
                 \Delta_\phi(p-l) \ T(s) ~.
\eeqa
By employing the Ward-Takahashi identities
\beq
k_\mu (2l+k)^{\mu} = \Delta_{\phi/\psi}^{-1}(l+k) - \Delta_{\phi/\psi}^{-1}(l)
\eeq
it is straightforward to show that
\beq
k_\mu \left(T_5^\mu +T_6^\mu  \right) = (e_\phi + e_\psi) \ T(s')  \ [G(s) - G(s')] \  T(s) \ .
\eeq
The last expression can be rewritten by making use of the BSE
\beq
T(s')[G(s)-G(s')]T(s) = T(s')\, [g^{-1}T(s)-1]  - [T(s')g^{-1}-1] \, T(s) = T(s)-T(s') .
\eeq
Adding up all contributions we arrive at
\beq
k_\mu \sum_{i=1}^{6} T_i^\mu =0 ~,
\eeq
which confirms the gauge invariance of the photon coupling to the bubble chain.

At the same time, insertion of the photon coupling at all possible places in the bubble chain
guarantees unitarity of the scattering matrix up to radiative corrections of order ${\cal O}(e^3)$.
To this end, we remark that
in the transition $\phi \psi \gamma \to  \phi \psi$ 
we can restrict ourselves to the states $|\phi \psi \rangle$ and
$|\phi \psi \gamma\rangle$. Unitarity of the ${\cal S}$-matrix,
${\cal S} = 1 -i {\cal T}$, implies then
\beq  \label{eq:unitar}
\langle \phi \psi | {\cal T} -{\cal T}^\dagger | \phi \psi \gamma \rangle
= - i \int_{PS} \Big\{ \langle \phi \psi | {\cal T}^\dagger | \phi' \psi' \rangle 
                  \langle \phi' \psi' | {\cal T} | \phi \psi \gamma \rangle  
              + \langle \phi \psi | {\cal T}^\dagger | \phi' \psi' \gamma' \rangle 
                  \langle \phi' \psi' \gamma' | {\cal T} | \phi \psi \gamma \rangle \Big\} \ ,
\eeq
where $\int_{PS} $ denotes the phase space integral for the set of intermediate states
$|\phi' \psi' \rangle$ and $|\phi' \psi' \gamma'\rangle$. We have introduced a superscript
for the intermediate particles with running three-momenta, $\phi'$, $\psi'$ and $\gamma'$,  
in order to distinguish them from the external particles $\phi$, $\psi$ and $\gamma$
with fixed momenta.
Note that the last matrix element $\langle \phi' \psi' \gamma' | {\cal T} | \phi \psi \gamma \rangle $
contains a disconnected piece of order ${\cal O}(e^0)$ in which the photon does not couple
to the bubble chain and thus appears in the ${\cal O}(e)$ part of the unitarity relation.
The remaining connected diagrams of this matrix element which contain the coupling of the
photon to the bubble chain are of order ${\cal O}(e^2)$ and will be neglected in the following.
By making use of the symmetry $\langle i | {\cal T}| j \rangle=
\langle j | {\cal T} | i \rangle$ due to time reversal invariance,
Eq.~(\ref{eq:unitar}) can be rewritten as
\beq  \label{cutko}
- 2 \Im \langle \phi \psi | {\cal T} | \phi \psi \gamma \rangle
=  \int_{PS} \Big\{ \langle \phi \psi | {\cal T}| \phi' \psi' \rangle^*
                  \langle \phi' \psi' | {\cal T} | \phi \psi \gamma \rangle  
              + \langle \phi \psi | {\cal T}| \phi' \psi' \gamma \rangle^* 
                  \langle \phi' \psi' \gamma | {\cal T} | \phi \psi \gamma \rangle \Big\} \ ,
\eeq
where now the phase space integral applies only to the particles $\phi'$ and $\psi'$.
The last equation represents the Cutkosky cutting rules \cite{C}. At the diagrammatic level
this amounts to cutting a pair
of $\phi$ and $\psi$ propagators at all possible places in the bubble chain
(keeping in mind that the photon is merely an external particle).
The two terms on the right hand side represent the two possibilities for the photon
to couple to the bubble chain before or after the cut.
Since these are the only possible cuts in the bubble chain
leading to imaginary pieces in the relevant kinematic region, one verifies Eq.~(\ref{cutko})
and hence unitarity of the ${\cal S}$-matrix up to radiative corrections.
However, if the photon couples to a propagator,
there is also the possibility to cut in the corresponding diagram
both propagators which are directly connected to the photon. For a photon with
$k^2 < 4 \, \mbox{min}  (m^2,M^2)$  which is the case both for physical photons
and for virtual photons from electron scattering, these cuts do not yield imaginary
values and can be safely omitted here.

Having convinced ourselves that it is possible to obtain gauge invariant and 
(up to radiative corrections) unitary
amplitudes by taking into account the coupling of the photon to scalar fields
in all possible ways in the bubble chain, 
we can now continue by applying this procedure to the slightly more
complicated case of the chiral effective meson-baryon Lagrangian.

%%%%%%%%%%%%%%%%%%%%%%%%%%%%%%%%%%%%%%%%%%%%%%%%%%%%%%%%%%%%%%%%%%%%%%%%%%%%%%%%
\section{Weinberg-Tomozawa term} \label{sec:wt}

The chiral effective Lagrangian describing the interactions between the
octet of Goldstone bosons $(\pi, K, \eta)$ and the ground
state baryon octet $(N, \Lambda, \Sigma, \Xi)$
is given at leading chiral order by
\beq \label{LagrphiB1}
\Lagr_{\phi B}^{(1)}  =  i \langle \bar{B} \gamma_{\mu} [D^{\mu},B] \rangle
                           - m_0 \langle \bar{B}B \rangle  + \ldots \ ,
\eeq
where $\langle \ldots \rangle$ denotes the trace in flavor space.
The $3 \times 3$ matrix $B$ collects the ground state baryon octet and 
$m_0$ is the common baryon octet mass in the chiral limit. We have only
displayed the terms relevant for the present investigation and omitted two operators which
contain the axial vector couplings of the mesons to the baryons. 
In the present investigation, we restrict ourselves to interaction kernels of the BSE
given by contact interactions. The axial vector couplings of the mesons to the baryons
could in principle contribute via direct and crossed Born terms, but the crossed
Born term corresponds to three-body intermediate states which
are beyond the scope of this work and we neglect the Born terms
throughout. (The inclusion of Born terms in the interaction kernel is deferred to future work
\cite{BBMN}.) 
In fact, many coupled-channels approaches only
take into account the contact interaction originating 
from the Lagrangian in (\ref{LagrphiB1}), see e.g. \cite{OOR} and references therein.

The covariant derivative of the baryon field is given by
\beq \label{CoDer}
[D_\mu, B] = \partial_\mu B + [ \Gamma_\mu, B]
\eeq
with the chiral connection
\beq
\label{Conn}
\Gamma_\mu = \frac{1}{2} [ u^\dagger,  \partial_\mu u] - \frac{i}{2} \Big( 
       u^\dagger v_\mu u + u v_\mu u^\dagger \Big)~, 
\eeq
and $v_\mu = - e {\cal A}_\mu Q$, where $Q= \frac{1}{3}\mbox{diag}(2, -1, -1)$ 
is the quark charge matrix. The pseudoscalar meson octet $\phi$ is arranged in a matrix valued field
\beq \label{Uphi}
U(\phi) = u^2(\phi) = \exp{\left( \sqrt{2} i \frac{\phi}{F} \right)} ~,
\eeq
with $F$  the pseudoscalar decay constant in the chiral limit.
Expansion of the chiral connection in the meson fields $\phi$
yields at leading order a $\phi^2 \bar{B} B$ contact interaction, the Weinberg-Tomozawa term,
which we choose to be the driving term for the BSE in this section.

The mesonic piece of the Lagrangian at leading chiral order is given by  \cite{GL}
\beq \label{Lagrphi} 
\Lagr_\phi^{(2)} =  \frac{F^2}{4} \langle \nabla_{\mu}U^\dagger \nabla^{\mu}U  \rangle 
            + \frac{F^2}{4} \langle \chi_+ \rangle , 
\eeq
where $\chi_+ = 2 B_0 (u^\dagger \mathcal{M} u^\dagger + u \mathcal{M} u)$ 
describes explicit chiral symmetry breaking via the quark mass
matrix $\mathcal{M} = \diag{(m_u, m_d, m_s)}$, and 
$B_0 = - \bra{0} \bar{q} q \ket{0} / F^2$ represents the order parameter of 
spontaneously broken chiral symmetry. 
The covariant derivative of the meson fields is given by (neglecting external axial-vector fields)
\beq
\nabla_{\mu}U = \partial_\mu U - i v_\mu U + i U v_\mu \ .
\eeq
In the Bethe-Salpeter formalism we choose to work with the propagators
\beqa
\Delta_i(p) &=& \frac{1}{p^2-M_i^2} ~, \no \\
S_a(p) &=& \frac{1}{\slashed{p}-m_a} ~,
\eeqa
with flavor indices $i,a$ and physical meson and baryon masses, $M_i$ and
$m_a$, respectively.
However, the following calculations are valid for all propagators
satisfying the Ward-Takahashi identities
\beqa
k^\mu V_\mu^\phi (p+k, k) = \Delta^{-1} (p+k) - \Delta^{-1} (p)\, , \no \\
k^\mu V_\mu^B (p+k, k) = S^{-1} (p+k) - S^{-1} (p) \ ,
\eeqa
where $V_\mu^{\phi} \, (V_\mu^{B} )$ are the corresponding 
$\gamma \phi^2 \, (\gamma \bar{B}B)$ three-point functions.

In the presence of a general interaction kernel $A$ and coupled channels
consisting of a meson-baryon pair the BSE for the process $\phi(q_i) B(p_i) \to \phi(q_f) B(p_f)$ 
generalizes to 
\beqa \label{BSEgen}
T_{fi} (p; q_f, q_i) &=& A_{fi} (p; q_f, q_i) +  i \sum_l \int_k 
  T_{fl} (p; q_f, k) \, \Delta_j(k) \, S_a(p+k) \ A_{li} (p; k, q_i)  \no \\
&=& A_{fi} (p; q_f, q_i) +  i \sum_l \int_k 
  A_{fl} (p; q_f, k) \, \Delta_j(k) \, S_a(p+k) \ T_{li} (p; k, q_i)
\eeqa
with $p = p_i +q_i= p_f+q_f$ and $l=\{\phi_j,B_a\}$ the channels which couple both to the initial
and final state, $i$ and  $f$. Note that we have replaced the common mass of the ground state baryon
octet, $m_0$, by the physical baryon masses $m_a$. This is consistent with the chiral order
of the interaction kernel derived from the Weinberg-Tomozawa term and, in particular, produces 
the unitarity cuts at the physical thresholds.

\begin{figure}[htb]
\centering
\includegraphics[width=0.6\textwidth]{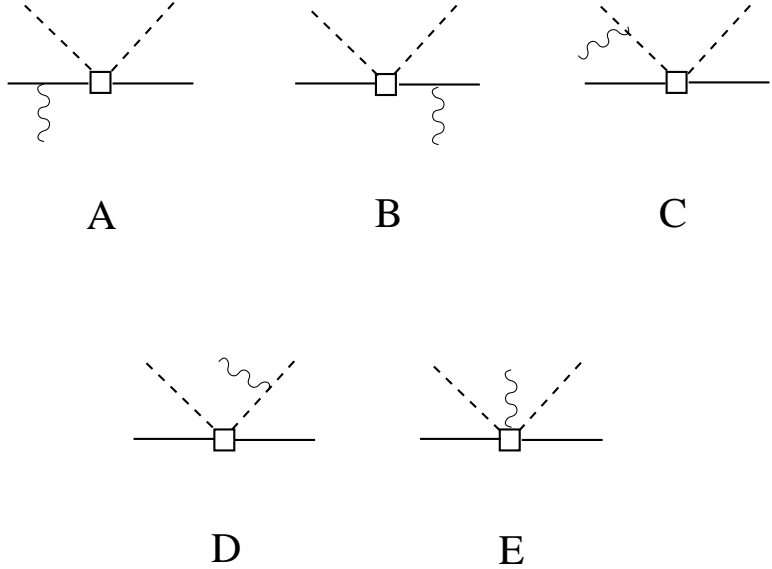} 
\caption{Tree diagrams for the process $\gamma \phi B \to \phi B$.
Solid, dashed and wavy lines correspond to baryons, mesons and photons, respectively.
The square denotes the vertex from the leading order Lagrangian.}
\label{fig:tree}
\end{figure}

After setting up the formalism we first calculate the tree level contributions
for the coupling of a photon to meson-baryon scattering. The pertinent Feynman diagrams
for the process $\gamma(k) \phi_i(q_i) B_a(p_i) \to \phi_j(q_f) B_b(p_f)$
are depicted in Figure~\ref{fig:tree}. In addition to the coupling of the photon to the
propagators the chiral connection in Eq.~(\ref{LagrphiB1}) gives rise to a 
$\gamma \phi^2 \bar{B}B$ vertex, Fig.~\ref{fig:tree}~E.

The tree contributions 
to the transition amplitude read
\beqa \label{photontree}
T^{\textit{(tree)} \, {bj,ai} }_\mu &=& - \frac{e}{4 F^2} \, \bigg\{ 
             (\slashed{q}_i + \slashed{q}_f) \, S_a(p_i+k)  \, \gamma_\mu \hat{Q}^a  
               + \gamma_\mu \hat{Q}^b S_b(p_f-k) \, (\slashed{q}_i + \slashed{q}_f) \no \\
 &&  \qquad + (\slashed{q}_i + \slashed{q}_f + \slashed{k} ) \, \Delta_i(q_i+k)  \,  
               [2q_i +k]_\mu \hat{Q}^i  \no \\
 &&  \qquad + (\slashed{q}_i + \slashed{q}_f - \slashed{k} ) \, \Delta_j(q_f-k)  \,  
                [2q_f -k]_\mu \hat{Q}^j    \no \\
 &&  \qquad -  \gamma_\mu  \big( \hat{Q}^j +\hat{Q}^i \big)\bigg\} \
       \big\langle \lambda^{b \dagger} [[ \lambda^{j \dagger}, \lambda^i ], \lambda^a ] \big\rangle  \ ,
\eeqa
where $\hat{Q}^a \lambda^a = [Q, \lambda^a]$ (no summation over $a$)
is the charge of the particle $a$ in units of $e$
and the $\lambda^i$ are the generators of the SU(3) Lie-Algebra in the physical basis.
By multiplying the tree contributions in Eq.~(\ref{photontree}) with $k_\mu$ 
gauge invariance is easily verified, if the momenta of the particles are put on-shell.

\begin{figure}[htb]
\centering
\includegraphics[width=0.7\textwidth]{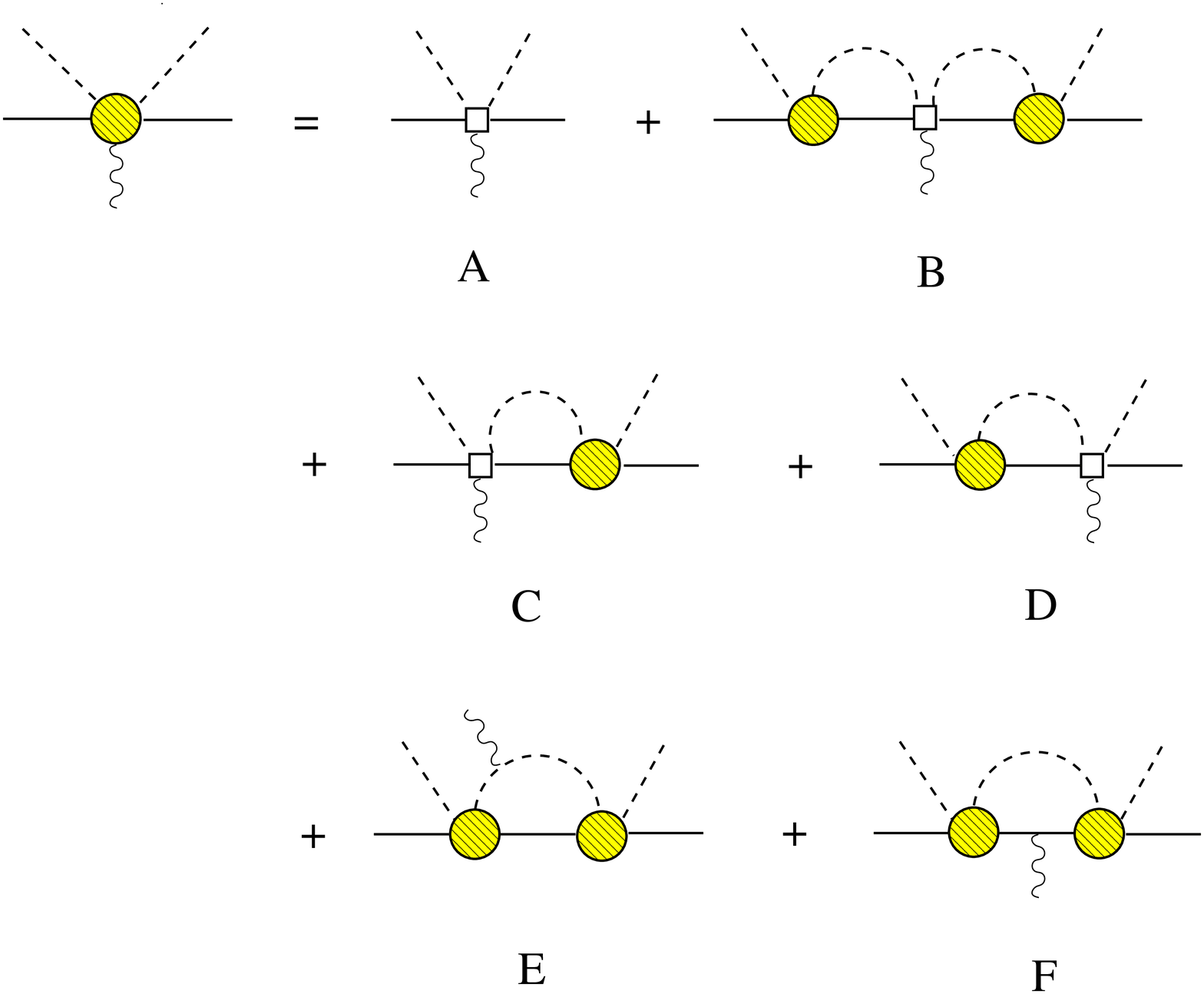} 
\caption{Bubble chain diagrams for the process $\gamma \phi B \to \phi B$.
Solid, dashed and wavy lines correspond to baryons, mesons and photons, respectively.
The square denotes the vertex from the leading order Lagrangian, the filled circle represents
the bubble chain derived from the BSE.}
\label{fig:bubblechain}
\end{figure}

In order to prove gauge invariance for the coupling of the photon to the bubble chain,
it is convenient to consider first the diagrams presented
in Fig.~\ref{fig:bubblechain} with the pertinent contributions given by
($a,b,c,d$ denote baryon flavor indices, whereas $i,j,m,n$ represent meson flavors):\\

\noindent
Fig.~\ref{fig:bubblechain}A:
\beq  \label{bubblefirst}
 A_\mu^{bj,ai}  = \frac{e}{4 F^2} \,  \gamma_\mu  \big( \hat{Q}^j +\hat{Q}^i \big) \ 
       \big\langle \lambda^{b \dagger} [[ \lambda^{j \dagger}, \lambda^i ],
       \lambda^a ] 
\big\rangle~,
\eeq

\noindent
Fig.~\ref{fig:bubblechain}B:
\beq
i^2 \int_l \int_q \  T^{bj,dn} (p';q_f,q) \  S_d (p'-q)  \Delta_n(q)
  \ A_\mu^{dn,cm} S_c (p-l)  \Delta_m(l)  \ T^{cm,ai} (p;l,q_i)~,
\eeq

\noindent
Fig.~\ref{fig:bubblechain}C:
\beq
i \int_l  \  T^{bj,cm} (p';q_f,l) \  S_c (p'-l)  \Delta_m(l) 
     \ A_\mu^{cm,ai}~,
\eeq

\noindent
Fig.~\ref{fig:bubblechain}D:
\beq
i \int_l \  A_\mu^{bj,cm} S_c (p-l)  \Delta_m(l)  \ T^{cm,ai} (p;l,q_i)~,
\eeq

\noindent
Fig.~\ref{fig:bubblechain}E:
\beq
i^2 \int_l \  T^{bj,cm} (p';q_f,l+k) \  S_c (p-l)  \Delta_m(l+k) 
     \ (-i e \hat{Q}^m (2l+k)_\mu) \Delta_m(l)  \ T^{cm,ai} (p;l,q_i)~,
\eeq

\noindent
Fig.~\ref{fig:bubblechain}F:
\beq  \label{bubblelast}
i^2 \int_l \  T^{bj,cm} (p';q_f,l) \   S_c (p'-l)
     \ (-i e \hat{Q}^c \gamma_\mu) S_c (p-l)  \Delta_m(l)  \ T^{cm,ai} (p;l,q_i)~,
\eeq
with $p' = p + k = p_i + q_i + k= p_f + q_f$.

For general momenta $p,q$ (i.e.\ not necessarily on-shell), the quantity 
$A_\mu^{bj,ai}$ satisfies the relation
\beq \label{relation}
k^\mu A_\mu^{bj,ai} = e \Bigl\{\big( \hat{Q}^b- \hat{Q}^a \big)  A^{bj,ai} (q, p) 
                 - \hat{Q}^i A^{bj,ai} (q, p+k) + \hat{Q}^j A^{bj,ai} (q-k, p) \Bigr\} \ ,
\eeq
where $A$ is the amplitude deduced from the Weinberg-Tomozawa term
\beq
A^{bj,ai} (q,p) = - \frac{1}{4 F^2} \,  (\slashed{q} + \slashed{p}) 
       \big\langle \lambda^{b \dagger} [[ \lambda^{j \dagger}, \lambda^i ], \lambda^a ] \big\rangle  \ .
\eeq

Making extensively use of Eq.~(\ref{relation}) and the BSE it is straightforward to show that
the contributions of Eqs.~(\ref{bubblefirst})-(\ref{bubblelast}) multiplied
by $k_\mu$ yield in total
\beq
e \bigl\{- \hat{Q}^a   T^{bj,ai} (p'; q_f, q_i) + \hat{Q}^b T^{bj,ai} (p; q_f, q_i)  
                      - \hat{Q}^i T^{bj,ai} (p';q_f, q_i+k) + \hat{Q}^j T^{bj,ai} (p;q_f-k, q_i)\bigr\} \ .
\eeq
This compensates exactly the contributions from the remaining four diagrams where the photon couples to
the external on-shell legs of the bubble chain. 
We have thus confirmed that gauge invariance is achieved if all possible
diagrams of a photon coupling to a bubble chain are taken into account. In particular,
it is not sufficient to consider only the coupling of the photon to external legs,
since this will lead to a gauge dependent amplitude.
On the other hand, within the field theoretical framework applied here it is also
not sufficient to take into account only the coupling of the photon to the interaction kernel.
Note that in the proof given here we do not assume the so-called on-shell
approximation for the interaction kernel.

It is also important to stress that an explicit evaluation of the BSE 
was not necessary and thus we do not need to specify the 
regularization scheme to render the loop integral in the BSE finite. 
Any regularization procedure which satisfies the Ward identities 
both for the propagators, Eq.(23), and the interaction kernel, 
Eq.(34), will maintain gauge invariance in the BSE as outlined in 
the proof.

Finally, the unitarity constraint for the full amplitude $T_\mu$ can be proven in analogy to the
treatment in Sec.~\ref{sec:complex}. The detailed derivation of unitarity is deferred to
Sec.~\ref{sec:uni}.
Thus, also in order to obtain
a unitarized amplitude (up to radiative corrections) one must take into account
the coupling of the photon at all possible places in the bubble chain.

%%%%%%%%%%%%%%%%%%%%%%%%%%%%%%%%%%%%%%%%%%%%%%%%%%%%%%%%%%%%%%%%%%%%%%%%%%%%%%%%
\section{Higher order interaction kernels}  \label{sec:nlo}

In this section, we will consider more complicated structures for the interaction
kernel as they arise at higher chiral orders in the effective Lagrangian.
Since we restrict ourselves to $\phi^2 \bar{B} B$ contact interactions, the most
general form of a term without the chiral invariant field strength tensor 
$f^+_{\mu\nu}$
is given by 
\beq  \label{genstruct}
\Lagr_{\textit{int}} = \bar{B} \ C_{\mu_1 \cdots \mu_l \mu_{l+1}  \cdots \mu_m \mu_{m+1}  \cdots \mu_n } 
   \Big( D^{\mu_1} \ldots D^{\mu_l} \tilde{\phi}\Big) \Big(D^{\mu_{l+1}} \ldots D^{\mu_m} \phi^\dagger\Big)
           \Big(    D^{\mu_{m+1}} \ldots D^{\mu_n} \tilde{B} \Big) \ ,
\eeq
where we have suppressed flavor indices for brevity (we merely kept  the symbol 
`` $\tilde{}$ '' as a reminder
of the flavor structure indicating that the in- and outgoing baryons and mesons can be different) 
and introduced the notation $D_{\mu} = \partial_\mu + i \hat{e} {\cal A}_\mu$
with $\hat{e}$ the charge of the particle $D_{\mu}$ is acting on.
As the contact interactions originate from a gauge invariant Lagrangian, charge
conservation is guaranteed at each vertex.
The introduction of explicit flavor indices does not change any of the following
conclusions. Note that the constant $C$ in Eq.~(\ref{genstruct}) may also contain elements of the
Clifford algebra.
From this Lagrangian one derives both a $ \tilde{\phi} (\tilde{q})  \tilde{B}(\tilde{p})  
\to \phi  (q) B (p)$ 
contact interaction $A(\tilde{p},\tilde{q},q)$ and a
$ \gamma (k) \tilde{\phi} (\tilde{q})\tilde{B}(\tilde{p}) \to  \phi (q)
B (p)  $ vertex $\epsilon_\mu A^\mu(\tilde{p},\tilde{q},q,k)$, 
where $\epsilon_\mu$ is the polarization vector of the photon.
Due to the form of the contact term (\ref{genstruct}), which can always be obtained
by partial integration, the vertices do not depend explicitly on the momentum $p$.
In Appendix~\ref{app} it is shown that
they satisfy the relation
\beqa \label{relationtwo}
k_\mu A^\mu (p_i,q_i,q_f,k) &=& \hat{e}_\phi A (p_i, q_i, q_f -k) 
                  - \hat{e}_{\tilde{\phi}} A (p_i, q_i +k, q_f)  \no \\
  &+& \hat{e}_B A (p_i, q_i, q_f) -\hat{e}_{\tilde{B}} A (p_i+k, q_i , q_f)
\eeqa
for general momenta of the particles. This equation is the analog
of Eq.~(\ref{relation}) for the Weinberg-Tomozawa term.
It follows then immediately by applying the same arguments as in the previous section
that the coupling of the photon to the two-particle state of the 
BSE yields a gauge invariant amplitude
also in the presence of more complicated contact interactions of the 
type Eq.~(\ref{genstruct}). 

Note that throughout we have not considered the dimension two magnetic moment
coupling $\sim \sigma^{\mu\nu} f^+_{\mu\nu}$ and higher order operators
involving the chiral covariant field strength tensor
$f^+_{\mu\nu}$. Such terms are of course present in the effective Lagrangian and must be considered at the
appropriate order in the chiral expansion of the interaction kernel. However, these 
are of the form $\partial_\nu v_\mu ( {\cal O}^{\nu \mu} - {\cal O}^{\mu \nu})$ with
some operator ${\cal O}^{\nu \mu}$ and the pertinent vertex in momentum space
vanishes upon contraction with the photon momentum
$k_\mu$.

As already mentioned in the previous section, the proof of gauge invariance does
not depend on the specific choice of the meson and baryon propagators, but is rather
valid for all propagators satisfying the Ward-Takahashi identities with the corresponding
$\gamma \phi^2$ and $\gamma \bar{B} B$ three-point functions.
For example, one can define the BSE by employing propagators with the physical masses
for the intermediate states and derive the interaction kernel from the effective
Lagrangian to a given chiral order. This automatically produces the correct
physical thresholds of the unitarity cuts and we have followed this path
in the present investigation.

Alternatively, one may prefer to deduce the propagators from the effective Lagrangian
as well. To leading chiral order this implies a common baryon octet mass shifting 
the threshold of the unitarity cuts to unphysical values. At higher chiral orders
the inclusion of self-energy diagrams for the meson and baryon propagators will
cure the situation by restoring the physical thresholds. The self-energy diagrams will modify
the simple form of the propagators given in  Eq.~(\ref{BSEgen}); e.g.,
the meson propagators will acquire the form
\beq
\Delta (p) = \frac{Z}{p^2 -M^2 - \Sigma_R (p)}
\eeq
with $M$ being the physical meson mass, $Z$ the appropriate wave function
renormalization constant, and $\Sigma_R (p)$ the renormalized self-energy.

In order to prove gauge invariance in the latter approach, one must also take into account
the coupling of the photon to the self-energy corrections of the propagators.
Since the effective Lagrangian is gauge invariant, the corresponding propagators and
three-point functions satisfy the pertinent Ward-Takahashi identities and the proof
is equivalent to the one given in the previous section.

We also would like to compare the present investigation with the work of \cite{KB1, KB2}. 
Although similar in spirit,
the authors study therein an integral equation for the two-body Green's function, 
whereas we prefer to work with an equation for the scattering amplitude. ``Gauging'' the
integral equation for the Green's function as outlined in \cite{KB1, KB2}, 
i.e. adding a vector index $\mu$ to all the terms of the equation
such that a linear equation in $\mu$-labeled quantities results, 
amounts to attaching an external photon everywhere including the external legs.
In our framework, the straightforward application of the gauging method to the integral equation
for the scattering amplitude fails,
as in this case
the procedure does not yield the contributions where the photon couples to the external legs. 
(Of course, one
could correct this by adding the missing contributions by hand.)
On the other hand, our approach is more convenient within the chiral effective
framework, as it allows a direct comparison with the scattering amplitude derived in the
perturbative scheme of ChPT. 
Moreover, the authors of \cite{KB1, KB2} restrict themselves merely to
one two-body channel, 
whereas in the present investigation this is generalized to several coupled channels. 
In contrast to \cite{KB1, KB2} we explicitly specify the interaction kernel
by deriving the vertices from
the chiral effective Lagrangian and utilizing them as interaction kernels
in the BSE.

%%%%%%%%%%%%%%%%%%%%%%%%%%%%%%%%%%%%%%%%%%%%%%%%%%%%%%%%%%%%%%%%%%%%%%%%%%%%%%%%
\section{Unitarity} \label{sec:uni}

After having constructed a gauge invariant amplitude for $B\phi\gamma\rightarrow B\phi$ 
with Weinberg-Tomozawa or more general contact interaction kernels, 
we would like to investigate unitarity of the obtained amplitude. 
The calculation presented in this section generalizes the findings for the
scalar field theory presented at the end of Sec.~\ref{sec:complex}, as one must take care of the
non-commutative nature of the matrix amplitudes due to the Clifford algebra and coupled channels.
Moreover, we do not assume symmetry of the transition amplitudes under exchange of initial and final
states.

In operator form the statement of a unitary scattering matrix amounts to
\beq \label{eq:uniop}
{\cal T} - {\cal T}^\dagger = - i {\cal T}^\dagger {\cal T} \ .
\eeq
For brevity we introduce a short-hand notation for the BSE, Eq.~(\ref{BSEgen}),
\beq
T(p) = A + \int T(p)G(p)A = A + \int AG(p)T(p) ,
\eeq
where $p$ is the external momentum and $G = iS\Delta$. 
The BSE for the meson-baryon scattering amplitude $T$ is easily 
transformed into the unitarity relation
\begin{equation}  \label{eq:unimod}
T - \bar T = \int \bar T (G-\bar G)T 
\end{equation}
with $\bar O \equiv \gamma_0 O^\dagger \gamma_0$, as both $T$ and $G$ are elements of the Clifford algebra.
Note that the adjoint $O^\dagger$ also implies taking the transposed matrix in channel space. 
We will see that the quantity $G-\bar G$ is equal to setting the intermediate
meson-baryon pairs on-shell in Eq.~(\ref{eq:unimod}). For invariant energies
below the lowest three-particle threshold Eq.~(\ref{eq:unimod}) is thus equivalent to the
unitarity constraint (\ref{eq:uniop}).

If $T$ is an analytic function, one can apply the residue theorem and rewrite the
difference $G-\bar G$ as 
\beq
i(S(p-l)\Delta(l)+\bar S(p-l)\bar\Delta(l)) \rightarrow 
i(-2\pi i)^{2}\delta_+(l^{2}-M^{2})\delta_+((p-l)^{2}-m^{2}) \ [\slashed{p} - \slashed{l} +m] ,
\eeq
where $l$ and $p$ are the loop and external momentum, respectively,
and $\delta_+(k^{2}-\mu^{2})= \delta(k^{2}-\mu^{2}) \theta (k_0)$. For a detailed derivation
of this replacement, see e.g. \cite{PS}.
The last equation can even be generalized to non-analytic $T$ which does not have
coinciding poles with $G$.
In particular, the above replacement is valid if $T$ describes the solution of the 
BSE with polynomial interaction kernels, as can be seen by insertion of its defining equation (\ref{BSEgen})
into Eq.~(\ref{eq:unimod}).

In order to prove unitarity for the transition $B\phi\gamma\rightarrow B\phi$, it is
convenient to introduce the amplitude
\beq
M^{\mu}_{\phi \gamma}  = V^{\mu}_{disc} + T^{\mu} 
\eeq
with the disconnected piece
\beq
V^{\mu}_{disc}  = 2E_{q_{i}}(2\pi)^{3}\delta^{(3)}(\mbox{\boldmath$q$}_{i}-
\mbox{\boldmath$q$}_{f})V^{\mu}_{B} 
+ 2E_{p_{i}}(2\pi)^{3}\delta^{(3)}(\mbox{\boldmath$p$}_{i}-
\mbox{\boldmath$p$}_{f})V^{\mu}_\phi 
\eeq
and $T^{\mu} $ the transition amplitude calculated in Sections \ref{sec:wt}
and \ref{sec:nlo}. The energies of the particles are given by
$E_{q_{i}} = \sqrt{q_{i}^{2} + M_{\phi}^{2}}$ and $E_{p_{i}}=\sqrt{p_{i}^{2}+m_{B}^{2}}$, respectively.
The piece $V^{\mu}_{disc}$ is represented by the two disconnected diagrams in which
the photon couples either to the baryon or meson, while the other particle
does not interact at all.
Although $V^{\mu}_{disc}$ does not 
contribute to on-shell matrix elements and could in principle be omitted
in the unitarity relation, its introduction generalizes unitarity beyond the physical region.
From the amplitude $M^{\mu}_{\phi \gamma}$ one constructs the reversed amplitude $M^{\mu}_{\gamma \phi}$
for the process $B(p_{f})\phi(q_{f})\rightarrow B(p_{i})\phi(q_{i})\gamma(k)$.
Neglecting radiative corrections and below the lowest three-particle (i.e.\ baryon two-meson)
threshold unitarity implies
\beq  \label{eq:muni}
M^{\mu}_{\phi \gamma}-\bar M^{\mu}_{\gamma \phi} = \int \bar T(p')(G(p')-\bar G(p'))M^{\mu}_{\phi \gamma} 
+ \int \bar M^{\mu}_{\gamma\phi }(G(p)-\bar G(p))T(p) \ ,
\eeq
where we have replaced again the two-body phase space integration by the four dimensional
integral over $G - \bar G$. Note that in the physical region $M^{\mu}_{\phi \gamma}$
reduces to $T^\mu$.
By inserting the amplitudes $T^\mu$ and $T$ from the BSE and making use of the
unitarity statement for $T$, Eq.~(\ref{eq:unimod}), as well as $\bar V^{\mu}_{\phi/B} = V^{\mu}_{\phi/B}$
and $\bar A^\mu = A^\mu$ from Eq.~(\ref{bubblefirst}), 
one can indeed confirm the unitarity constraint for
$M^{\mu}_{\phi \gamma}$ and thus for $T^\mu$ for on-shell matrix elements.

We refrain from presenting the entire and tedious calculation here, 
but would like to comment on two points.
First, the contribution of the disconnected graphs drops out on the l.h.s. of the 
unitarity statement (\ref{eq:muni}) (due to the symmetry of these graphs under interchange 
of incoming and outgoing particles and $\bar V^{\mu}_{\phi/B} = V^{\mu}_{\phi/B}$). 
On the r.h.s., they produce terms of the type
\beqa
& &\int_{l}\bar T(p') \ (G(p')-\bar G(p')) \ 2E_{q_{i}}(2\pi)^{3}\delta^{(3)}(\mbox{\boldmath$q$}_{i}-
\mbox{\boldmath$l$})V^{\mu}_{B}  \no \\
&=& \int_{l} \bar T(p')\ i(-2\pi i)^{2} \, \delta_+((p'-l)^{2}-m^{2}) \, \delta_+(l^{2}-M^{2}) \
 [\slashed{p}' - \slashed{l} +m] \ 2E_{q_{i}}(2\pi)^{3}\delta^{(3)}(\mbox{\boldmath$q$}_{i}-
\mbox{\boldmath$l$})V^{\mu}_{B}  \no \\
&=& \int d^{3}l \ \bar T(p')(-2\pi i) \ \delta_+((p'-l)^{2}-m^{2}) \  [\slashed{p}' - \slashed{l} +m] \
\delta^{(3)}(\mbox{\boldmath$q$}_{i}- \mbox{\boldmath$l$})V^{\mu}_{B} \no  \\
&=& \bar T(p')(-2\pi i) \ \delta_+((p'-q_{i})^{2}-m^{2})  [\slashed{p}' - \slashed{q}_i +m] \
V^{\mu}_{B} \no  \\[0.3cm]
&=& \bar T(p') \ [S(p_{i}+k) - \bar S(p_{i}+k)] \ V^{\mu}_{B} .
\eeqa
For on-shell matrix elements (and $k \ne 0$) this vanishes as $\epsilon \to 0$ in the propagators, 
but it happens
that all terms of this type cancel each other on the r.h.s. of Eq.~(\ref{eq:muni})
even for general external momenta.
Our second comment concerns some contributions from diagrams \ref{fig:bubblechain}E and F.
On the l.h.s. of Eq.~(\ref{eq:muni}) one obtains, e.g., the combination 
\beq
i\int_{l} \, \bar T(p') \, S(p'-l) \, \Delta(l) \, V^{\mu}_{B} \, S(p-l) \, T(p) 
+ i\int_{l} \, \bar T(p') \, \bar S(p'-l) \, \bar \Delta(l) \, V^{\mu}_{B} \, \bar S(p-l) \, T(p) 
\eeq
which represents the discontinuity of the transition amplitude stemming from
Fig.~\ref{fig:bubblechain}F (the two integrals only differ in the sign of
the ``$i \epsilon$'' terms in the propagators).
According to the Cutkosky cutting rules and for momenta $k^2 < 4 m^2$ 
this discontinuity is given by
\beq
\int_{l} \,  \bar T(p') \, (G(p')-\bar G(p')) \, V^{\mu}_{B} \, S(p-l) \, T(p) 
+ \int_{l} \, \bar T(p') \, \bar S(p'-l) \, V^{\mu}_{B} \, (G(p)-\bar G(p)) \, T(p) .
\eeq
In the kinematical region which is of relevance here the cut through the two
propagators connecting to the photon does not contribute to the discontinuity and can be
safely neglected.
We conclude by emphasizing that the unitarity constraint (\ref{eq:muni}) is only fulfilled 
if the photon couples to all possible places in the bubble chain.

%%%%%%%%%%%%%%%%%%%%%%%%%%%%%%%%%%%%%%%%%%%%%%%%%%%%%%%%%%%%%%%%%%%%%%%%%%%%%%%%
\section{Conclusions} \label{sec:conclusions}

In the present work, we have studied how gauge invariance is obtained for a photon
coupling to a two-body state described by the solution of the Bethe-Salpeter equation.
We have discussed the procedure both for a simple complex scalar field theory
and for interaction kernels derived from chiral effective Lagrangians in the meson-baryon sector. 
In the latter
case, we have first considered the Weinberg-Tomozawa term and afterwards the most general
contact interaction consisting of two mesons and two baryons which can arise
in the chiral effective framework. Our study underlines that it is {\it not} sufficient
to take into account only the coupling of the photon to the external legs as
has been done in many calculations based on
chiral unitary approaches, but one rather has to include all possible contributions
of the photon coupling to the vertices and intermediate states.
Neither is it sufficient to consider only the coupling of the photon to the interaction kernel.
At the same time, coupling  of the photon to the bubble chain
at all possible places is necessary, in order to guarantee a unitary scattering matrix
up to radiative corrections.

For the interaction kernels discussed in the present work we have shown explicitly
that gauge invariance is maintained in this manner.
It is accomplished without assuming the on-shell approximation for
the interaction kernel.
Moreover, the explicit evaluation of the loop 
integral in the Bethe-Salpeter equation is not necessary and hence 
the proof does not depend on the chosen regularization scheme. 
But the regularization procedure is required to satisfy the Ward identities both 
for the propagators and the interaction kernels.
This study will also be of importance for photo- and electroproduction processes
of mesons on nucleons and for radiative decays of baryons and mesons
which must be treated in a similar way, in order to achieve gauge invariant
and unitarized
amplitudes. Work along these lines is in progress.

%%%%%%%%%%%%%%%%%%%%%%%%%%%%%%%%%%%%%%%%%%%%%%%%%%%%%%%%%%%%%%%%%%%%%%%%%%%%%%%%
\section*{Acknowledgments}
We thank Bastian Kubis for discussions and reading the manuscript.
We thank Christoph Hanhart and Daniel Phillips for useful communications.
Partial financial support by Deutsche Forschungsgemeinschaft is gratefully acknowledged.
This research is part of the EU Integrated Infrastructure Initiative Hadronphysics under contract
 number RII3-CT-2004-506078.
 
%\vfill\eject 
 
%%%%%%%%%%%%%%%%%%%%%%%%%%%%%%%%%%%%%%%%%%%%%%%%%%%%%%%%%%%%%%%%%%%%%%%%%%%%%%%%
\appendix

%%%%%%%%%%%%%%%%%%%%%%%%%%%%%%%%%%%%%%%%%%%%%%%%%%%%%%%%%%%%%%%%%%%%%%%%%%%%%%%%
\section{Higher chiral orders}  \label{app}
%%%%%%%%%%%%%%%%%%%%%%%%%%%%%%%%%%%%%%%%%%%%%%%%%%%%%%%%%%%%%%%%%%%%%%%%%%%%%%%%

In this appendix we derive Eq.~(\ref{relationtwo}) which follows from
a contact interaction of the type 
 \beq  \label{app:lagr}
\Lagr_{\textit{int}} = \bar{B} \ C_{\mu_1 \cdots \mu_l \mu_{l+1}  \cdots \mu_m \mu_{m+1}  \cdots \mu_n } 
   \Big( D^{\mu_1} \ldots D^{\mu_l} \tilde{\phi}\Big) \Big(D^{\mu_{l+1}} \ldots D^{\mu_m} \phi^\dagger\Big)
           \Big(    D^{\mu_{m+1}} \ldots D^{\mu_n} \tilde{B} \Big)  
\eeq
with $D_{\mu} = \partial_\mu + i \hat{e} {\cal A}_\mu$.
The 
$ \gamma (k) \tilde{\phi} (q_i)  \tilde{B}(p_i) \to \phi (q_f) {B} (p_f) $ vertex 
is obtained from this Lagrangian by extracting the part linear in ${\cal A}_\mu$.
A partial derivative acting, e.g., on an incoming meson $\phi(q)$ yields a factor $-iq$ 
in momentum space, while a partial derivative on $\phi(q) {\cal A} (k)$ (both momenta incoming)
leads to $-i(q+k)$. Therefore, one obtains the vertex
\beqa
A_\mu &=&
i \sum_{s=1}^{l} C_{\mu_1 \cdots  \mu_n }\Big|_{\mu_s=\mu} \,(-i)^{(2l-2m+n)} 
  \left[\prod_{t=1}^{s-1} (q_i+k)^{\mu_t} \right] \bigl(-\hat{e}_{\tilde{\phi}}\bigr) 
  \left[\prod_{t=s+1}^{l} q_i^{\mu_{t}} \right]
  \left[\prod_{t=l+1}^{m} q_f^{\mu_{t}} \right]
  \left[\prod_{t=m+1}^{n} p_i^{\mu_{t}} \right] \no \\
%&& \qquad  \times \  i^{(m-l)} q_f^{\mu_{l+1}} \ldots q_f^{\mu_{m}} 
%                  (-i)^{(n-m)} p_i^{\mu_{m+1}} \ldots p_i^{\mu_{n}} \no \\
&+&
i \sum_{s=l+1}^{m} C_{\mu_1 \cdots  \mu_n }\Big|_{\mu_s=\mu} \,(-i)^{(2l-2m+n)} 
  \left[\prod_{t=1}^{l} q_i^{\mu_{t}} \right]
  \left[\prod_{t=l+1}^{s-1} (q_f-k)^{\mu_{t}} \right] \bigl(-\hat{e}_{\phi}\bigr)
  \left[\prod_{t=s+1}^{m} q_f^{\mu_{t}} \right]
  \left[\prod_{t=m+1}^{n} p_i^{\mu_{t}} \right] \no \\
%&& \qquad  \times \  i^{(m-l)} (q_f-k)^{\mu_{l+1}} \ldots (q_f-k)^{\mu_{s-1}}
%    (- \hat{e}_\phi) q_f^{\mu_{s+1}} \ldots q_f^{\mu_{m}} 
%                  (-i)^{(n-m)} p_i^{\mu_{m+1}} \ldots p_i^{\mu_{n}} \no \\
&+&
i \sum_{s=m+1}^{n} C_{\mu_1 \cdots  \mu_n }\Big|_{\mu_s=\mu} \,(-i)^{(2l-2m+n)} 
  \left[\prod_{t=1}^{l} q_i^{\mu_{t}} \right]
  \left[\prod_{t=l+1}^{m} q_f^{\mu_{t}} \right]
  \left[\prod_{t=m+1}^{s-1} (p_i+k)^{\mu_{t}} \right] \bigl(-\hat{e}_{\tilde{B}}\bigr)
  \left[\prod_{t=s+1}^{n} p_i^{\mu_{t}} \right] \no \\
\eeqa
Multiplying this equation with $k_\mu$ and making use of charge conservation at the vertex
which follows from gauge invariance of the Lagrangian one arrives at
\beqa
k^\mu A_\mu &=&
i C_{\mu_1 \cdots \mu_n } (-i)^{(2l-2m+n)} (- \hat{e}_{\tilde{\phi}})  (q_i+k)^{\mu_1} 
       \dotsm (q_i+k)^{\mu_{l}} 
       q_f^{\mu_{l+1}} \dotsm q_f^{\mu_{m}} p_i^{\mu_{m+1}} \dotsm p_i^{\mu_{n}} \no \\
&-&
i C_{\mu_1 \cdots \mu_n } (-i)^{(2l-2m+n)} (- \hat{e}_\phi) q_i^{\mu_1} \dotsm q_i^{\mu_{l}} 
       (q_f-k)^{\mu_{l+1}} \dotsm (q_f-k)^{\mu_{m}} p_i^{\mu_{m+1}} \dotsm p_i^{\mu_{n}} \no \\
&+&
i C_{\mu_1 \cdots \mu_n } (-i)^{(2l-2m+n)} (- \hat{e}_{\tilde{B}}) q_i^{\mu_1} \dotsm q_i^{\mu_{l}} 
       q_f^{\mu_{l+1}} \dotsm q_f^{\mu_{m}} (p_i+k)^{\mu_{m+1}} \dotsm (p_i+k)^{\mu_{n}} \no \\
&-&
i C_{\mu_1 \cdots \mu_n } (-i)^{(2l-2m+n)} (- \hat{e}_{B}) q_i^{\mu_1} \dotsm q_i^{\mu_{l}} 
       q_f^{\mu_{l+1}} \dotsm q_f^{\mu_{m}} p_i^{\mu_{m+1}} \dotsm p_i^{\mu_{n}} \ .
\eeqa
On the other hand, the vertex $\tilde{\phi} (q_i)  \tilde{B}(p_i) \to \phi (q_f) {B} (p_f) $
from the Lagrangian (\ref{app:lagr}) is given by the piece without the photon field ${\cal A}_\mu$
and reads
\beq
A(p_i, q_i, q_f) = 
  i C_{\mu_1 \cdots \mu_n } (-i)^{(2l-2m+n)} q_i^{\mu_1} \dotsm q_i^{\mu_{l}} 
       q_f^{\mu_{l+1}} \dotsm q_f^{\mu_{m}} p_i^{\mu_{m+1}} \dotsm p_i^{\mu_{n}} \ .
\eeq
This proves the Ward identity (\ref{relationtwo})
\beqa 
k^\mu A_\mu (p_i,q_i,q_f,k) &=& \hat{e}_\phi A (p_i, q_i, q_f -k) 
                  - \hat{e}_{\tilde{\phi}} A (p_i, q_i +k, q_f)  \no \\
  &+& \hat{e}_B A (p_i, q_i, q_f) -\hat{e}_{\tilde{B}} A (p_i+k, q_i , q_f)~.
\eeqa

%%%%%%%%%%%%%%%%%%%%%%%%%%%%%%%%%%%%%%%%%%%%%%%%%%%%%%%%%%%%%%%%%%%%%%%%%%%%%%%%
 
%\vfill\eject 


\begin{thebibliography}{99}

\bibitem{GL}
J.~Gasser and H.~Leutwyler,
%``Chiral Perturbation Theory: Expansions In The Mass Of The Strange Quark,''
Nucl.\ Phys.\ B {\bf 250} (1985) 465.
%%CITATION = NUPHA,B250,465;%%



\bibitem{KSW}
N.~Kaiser, P.~B.~Siegel and W.~Weise,
%``Chiral dynamics and the S11 (1535) nucleon resonance,''
Phys.\ Lett.\ B {\bf 362}, 23 (1995)
[arXiv:nucl-th/9507036];\\
%%CITATION = NUCL-TH 9507036;%%
N.~Kaiser, P.~B.~Siegel and W.~Weise,
%``Chiral dynamics and the low-energy kaon - nucleon interaction,''
Nucl.\ Phys.\ A {\bf 594}, 325 (1995)
[arXiv:nucl-th/9505043].
%%CITATION = NUCL-TH 9505043;%%




\bibitem{OM} 
J.~A.~Oller and U.-G.~Mei{\ss}ner,
%``Chiral dynamics in the presence of bound states: Kaon nucleon  interactions
%revisited,''
Phys.\ Lett.\ B {\bf 500}, 263 (2001)
[arXiv:hep-ph/0011146];
U.-G.~Mei{\ss}ner and J.~A.~Oller,
%``Chiral unitary meson baryon dynamics in the presence of resonances:  Elastic
%pion nucleon scattering,''
Nucl.\ Phys.\ A {\bf 673}, 311 (2000)
[arXiv:nucl-th/9912026].


\bibitem{LK} E.~E.~Kolomeitsev and M.~F.~M.~Lutz,
%``On baryon resonances and chiral symmetry,''
Phys.\ Lett.\ B {\bf 585} (2004) 243
[arXiv:nucl-th/0305101].
%%CITATION = NUCL-TH 0305101;%%

  
\bibitem{OO}
J.~A.~Oller and E.~Oset,
%``Chiral symmetry amplitudes in the S-wave isoscalar and isovector  channels
%and the sigma, f0(980), a0(980) scalar mesons,''
Nucl.\ Phys.\ A {\bf 620}, 438 (1997)
[Erratum-ibid.\ A {\bf 652}, 407 (1999)]
[arXiv:hep-ph/9702314];
%%CITATION = HEP-PH 9702314;%%
J.~A.~Oller, E.~Oset and J.~R.~Pelaez,
%``Meson meson and meson baryon interactions in a chiral non-perturbative
%approach,''
Phys.\ Rev.\ D {\bf 59}, 074001 (1999)
[Erratum-ibid.\ D {\bf 60}, 099906 (1999)]
[arXiv:hep-ph/9804209].
%%CITATION = HEP-PH 9804209;%%  
  
              
\bibitem{KWW}  
N.~Kaiser, T.~Waas and W.~Weise,
%``SU(3) chiral dynamics with coupled channels: Eta and kaon  photoproduction,''
Nucl.\ Phys.\ A {\bf 612}, 297 (1997)
[arXiv:hep-ph/9607459];
%%CITATION = HEP-PH 9607459;%%
J.~Caro Ramon, N.~Kaiser, S.~Wetzel and W.~Weise,
%``Chiral SU(3) dynamics with coupled channels: Inclusion of p-wave
%multipoles,''
Nucl.\ Phys.\ A {\bf 672}, 249 (2000)
[arXiv:nucl-th/9912053].
%%CITATION = NUCL-TH 9912053;%%



\bibitem{BWW} S.~D.~Bass, S.~Wetzel and W.~Weise,
%``Axial U(1) dynamics in eta and eta' photoproduction,''
Nucl.\ Phys.\ A {\bf 686}, 429 (2001)
[arXiv:hep-ph/0007293].
%%CITATION = HEP-PH 0007293;%%

\bibitem{BMW} 
B.~Borasoy, E.~Marco and S.~Wetzel,
%``eta , eta -prime photoproduction and electroproduction off nucleons,''
Phys.\ Rev.\ C {\bf 66}, 055208 (2002)
[arXiv:hep-ph/0212256].
%%CITATION = HEP-PH 0212256;%%

\bibitem{GR} 
F.~Gross and D.~O.~Riska,
%``Current Conservation And Interaction Currents In Relativistic Meson
%Theories,''
Phys.\ Rev.\ C {\bf 36}, 1928 (1987).
%%CITATION = PHRVA,C36,1928;%%


\bibitem{SG}  
Y.~Surya and F.~Gross,
%``Unitary, gauge invariant, relativistic resonance model for pion
%photoproduction,''
Phys.\ Rev.\ C {\bf 53}, 2422 (1996).
%%CITATION = PHRVA,C53,2422;%%


%\cite{Kvinikhidze:1998xn}
\bibitem{KB1}
A.~N.~Kvinikhidze and B.~Blankleider,
%``Gauging of equations method. I. Electromagnetic currents of three
%distinguishable particles,''
Phys.\ Rev.\ C {\bf 60} (1999) 044003
[arXiv:nucl-th/9901001].
%%CITATION = NUCL-TH 9901001;%%


%\cite{Kvinikhidze:1999xp}
\bibitem{KB2}
A.~N.~Kvinikhidze and B.~Blankleider,
%``Gauging of equations method. II. Electromagnetic currents of three identical
%particles,''
Phys.\ Rev.\ C {\bf 60} (1999) 044004
[arXiv:nucl-th/9901002].
%%CITATION = NUCL-TH 9901002;%%


%\cite{vanAntwerpen:1994vh}
\bibitem{AA}
C.~H.~M.~van Antwerpen and I.~R.~Afnan,
%``A Gauge invariant unitary theory for pion photoproduction,''
Phys.\ Rev.\ C {\bf 52} (1995) 554
[arXiv:nucl-th/9407038].
%%CITATION = NUCL-TH 9407038;%%


%\cite{Haberzettl:1997jg}
\bibitem{H}
H.~Haberzettl,
%``Gauge-invariant theory of pion photoproduction with dressed hadrons,''
Phys.\ Rev.\ C {\bf 56} (1997) 2041
[arXiv:nucl-th/9704057].
%%CITATION = NUCL-TH 9704057;%%


\bibitem{Ol} J.~A.~Oller,
%``The Phi $\to$ gamma K0 anti-K0 decay,''
Phys.\ Lett.\ B {\bf 426}, 7 (1998)
[arXiv:hep-ph/9803214].
%%CITATION = HEP-PH 9803214;%%

\bibitem{MHOT} E.~Marco, S.~Hirenzaki, E.~Oset and H.~Toki,
%``Radiative decay of rho0 and Phi mesons in a chiral unitary approach,''
Phys.\ Lett.\ B {\bf 470}, 20 (1999)
[arXiv:hep-ph/9903217].
%%CITATION = HEP-PH 9903217;%%

\bibitem{BN1} B.~Borasoy and R.~Ni{\ss}ler,
%``Two-photon decays of pi0, eta and eta',''
Eur.\ Phys.\ J.\ A {\bf 19}, 367 (2004)
[arXiv:hep-ph/0309011].
%%CITATION = HEP-PH 0309011;%%

\bibitem{BN2} B.~Borasoy and R.~Ni{\ss}ler,
%``eta, eta' $\to$ pi+ pi- gamma with coupled channels,''
Nucl.\ Phys.\ A {\bf 740}, 362 (2004)
[arXiv:hep-ph/0405039].
%%CITATION = HEP-PH 0405039;%%

\bibitem{HN} H. ~Haberzettl and K.~Nakayama, {\it in preparation}.

\bibitem{C} R.~E.~Cutkosky,
%``Solutions Of A Bethe-Salpeter Equations,''
Phys.\ Rev.\  {\bf 96}, 1135 (1954).
%%CITATION = PHRVA,96,1135;%%

\bibitem{BBMN} B.~Borasoy, P.~Bruns, U.-G.~Mei{\ss}ner and R.~Ni{\ss}ler, {\it in preparation}.

\bibitem{OOR} J.~A.~Oller, E.~Oset and A.~Ramos,
%``Chiral unitary approach to meson meson and meson baryon interactions  and
%nuclear applications,''
Prog.\ Part.\ Nucl.\ Phys.\  {\bf 45}, 157 (2000)
[arXiv:hep-ph/0002193].
%%CITATION = HEP-PH 0002193;%%



\bibitem{PS} M.~E.~Peskin and D.~V.~Schroeder, ``An Introduction to Quantum Field Theory'', 
            Westview Press (1995).





\end{thebibliography}
\end{document}